\documentstyle[preprint,prb,aps,floats,psfig]{revtex}

\begin{document}
\draft


\title{The Aharonov-Bohm effect for an exciton}

\author{R.~A.~R\"omer$^1$ and M.~E.~Raikh$^2$}

\address {$^1$Institut f\"ur Physik, Technische Universit\"at, D-09107
  Chemnitz, Germany\\ $^2$Department of Physics, University of Utah,
  Salt Lake City, Utah 84112, U.S.A.}


\maketitle

\begin{abstract}
  We study theoretically the exciton absorption on a ring threaded by
  a magnetic flux. For the case when the attraction between electron
  and hole is short-ranged we get an exact solution of the problem. We
  demonstrate that, despite the electrical neutrality of the exciton,
  both the spectral position of the exciton peak in the absorption,
  and the corresponding oscillator strength oscillate with magnetic
  flux with a period $\Phi_0$---the universal flux quantum. The origin
  of the effect is the finite probability for electron and hole,
  created by a photon at the same point, to tunnel in the opposite
  directions and meet each other on the opposite side of the ring.
\end{abstract}


\narrowtext


One of the manifestations of the Aharonov-Bohm (AB) effect
\cite{AhaB59} in the ring geometry \cite{ByeY61,ButIL83} is the
periodic dependence of the transmission coefficient for an electron
traversing the ring on the magnetic flux $\Phi$ through the
ring.\cite{GefIA84,ButILP85} The period of oscillations is equal to
$\Phi_0=hc/e$ --- the universal flux quantum.

For one-dimensional (1D) continuum interacting quantum systems with
translational invariance there is also a periodicity of many-particle
states as a functions of flux.\cite{SutS90,ShaS90,MulWL93,Kus94a}  In
1D lattice systems, the lifting of Galilean invariance allows for
various periodicities of the states.\cite{SutS90,ShaS90} For the
ground state, this behavior can be interpreted, according to the above
definition of $\Phi_0$, as a signature of the existence of elementary
excitations with multiple --- sometimes even fractional --- charges
.\cite{SutS90,RomS94b,RomS94c,Kus95,KriSSJ96}  In the case of strong
electron-electron interaction the adequate description of the
many-body states is based on excitations of the Wigner-crystal
.\cite{Los92,KriSSG95}  Furthermore, the absence of sensitivity to the
flux in such systems is an indication of the onset of the Mott
transition. \cite{ShaS90,Koh64,RomP95}  Similarly, the sensitivity of
single-particle energies to the flux \cite{Tho74} can be used as a
criterion of the Anderson-type metal-insulator transition in
disordered systems.\cite{And58}  Combined effects of interactions and
disorder in 1D have received much attention in the last decade
.\cite{RomP95,Dor90,She94,LeaRS99,RomLS99b}  Numerical studies of
pairing effects for two particles with repulsive interaction in a
disordered environment were carried out using the AB setting
.\cite{WeiMPF95}  Other physical manifestations of the AB effect in
the ring geometry considered in the literature include the evolution
of electron states for a time-dependent flux,\cite{GorKGS97} and a
flux-dependent equilibrium distortion of the lattice caused by
electron-phonon interactions. \cite{Kus92}

The physical origin of the flux sensitivity of an electron on the ring
is its charge which couples to the vector potential.  Correspondingly,
the coupling to the flux has the opposite sign for an electron and a
hole.  For this reason an {\em exciton}, being a bound state of
electron and hole and thus a {\em neutral}\ entity, should not be
sensitive to the flux.  However, due to the finite size of the
exciton, such a sensitivity will emerge. This effect is demonstrated
in the present paper.
Below we study the AB-oscillations both in the binding energy and in
the oscillator strength of the exciton absorption. We choose as a
model a short-range attraction potential between electron and hole,
which allows to solve the three-body problem (electron, hole, and a
ring) exactly. From this exact solution, we trace the behavior of the
AB oscillations when increasing the radius of the ring or the strength
of the electron-hole attraction.


Denote with $\varphi_e$ and $\varphi_h$ the azimuthal coordinates of
the electron and hole, respectively. In the absence of interaction the
wave functions of electrons and holes are given by
\begin{equation}
\label{eq-eigen}
\Psi_N^{(e)}(\varphi_e) =\frac{1}{\sqrt{2\pi}}e^{iN\varphi_e}, \quad
\Psi_{N^{\prime}}^{(h)}(\varphi_h)
=\frac{1}{\sqrt{2\pi}}e^{iN^{\prime}\varphi_h},
\end{equation}
where $N$ and $N^{\prime}$ are integers. The corresponding energies
are
\begin{equation}
\label{eq-energies}
E_N^{(e)}=\frac{\hbar^2}{2m_e\rho^2}\Biggl(N-\frac{\Phi}{\Phi_0}\Biggr)^2,
\quad E_{N^{\prime}}^{(h)}
=\frac{\hbar^2}{2m_h\rho^2}\Biggl(N^{\prime}+\frac{\Phi}{\Phi_0}
\Biggr)^2.
\end{equation}
Here $\rho$ is the radius of the ring, and $m_e$, $m_h$ stand for the
effective masses of electron and hole, respectively. In the presence
of an interaction $V\Bigl[R(\varphi_e-\varphi_h)\Bigr]$, where
$R(\varphi_e-\varphi_h)=2\rho\sin(\frac{\varphi_e-\varphi_h}{2})$ is
the distance between electron and hole, we search for the wave
function of the exciton in the form
\begin{equation}
\label{eq-wave}
\Psi(\varphi_e,\varphi_h)= \sum_{N,N^{\prime}}A_{N,N^{\prime}}
\Psi_N^{(e)}(\varphi_e)\Psi_{N^{\prime}}^{(h)}(\varphi_h).
\end{equation}
The coefficients $A_{N,N^{\prime}}$ are to be found from the equation
\begin{equation}
\label{eq-A}
\sum_{N,N^{\prime}}A_{N,N^{\prime}}
\Bigl[E_N^{(e)}+E_{N^{\prime}}^{(h)} - \Delta\Bigr]
\Psi_N^{(e)}(\varphi_e)\Psi_{N^{\prime}}^{(h)}(\varphi_h) +
V\Bigl[R(\varphi_e-\varphi_h)\Bigr]\Psi(\varphi_e,\varphi_h)=0,
\end{equation}
where $\Delta$ is the energy of the exciton. The formal expression for
$A_{N,N^{\prime}}$ follows from Eq.\ (\ref{eq-A}) after multiplying it
by
$\Bigl[\Psi_N^{(e)}(\varphi_e)
\Psi_{N^{\prime}}^{(h)}(\varphi_h)\Bigr]^{\dagger}$
and integrating over $\varphi_e$ and $\varphi_h$
\begin{equation}
\label{eq-formal}
A_{N,N^{\prime}}=
-\frac{1}{2\pi}\int_0^{2\pi}d\varphi_e\int_0^{2\pi}d\varphi_h
\frac{V\Bigl[R(\varphi_e-\varphi_h)\Bigr]\Psi(\varphi_e,\varphi_h)}
{E_N^{(e)}+E_{N^{\prime}}^{(h)} - \Delta}
e^{-i(N\varphi_e+N^{\prime}\varphi_h)}.
\end{equation}
At this point we make use of the assumption that the potential
$V\Bigl[R(\varphi_e-\varphi_h)\Bigr]$ is short-ranged. This implies
that the integral over $\varphi_h$ is determined by a narrow interval
of $\varphi_h$ close to $\varphi_e$.  Then we can replace $\varphi_h$
by $\varphi_e$ in the rest of the integrand. As a result, Eq.\
(\ref{eq-formal}) simplifies to
\begin{equation}
\label{eq-simplified}
A_{N,N^{\prime}}= -\frac{V_0}{E_N^{(e)}+E_{N^{\prime}}^{(h)} - \Delta}
\int_0^{2\pi}d\varphi_e\Psi(\varphi_e,\varphi_e)e^{-i(N+N^{\prime})\varphi_e},
\end{equation}
where the constant $V_0<0$ is defined as
\begin{equation}
\label{eq-not}
V_0=\frac{1}{2\pi}\int d\varphi V\Bigl[R(\varphi)\Bigr].
\end{equation}
Finally we derive a closed equation, which determines the exciton
energies.  This equation follows from Eqs.\ (\ref{eq-wave}) and
(\ref{eq-simplified}) as a self-consistency condition. Indeed, by
setting in Eq.\ (\ref{eq-wave}) $\varphi_e=\varphi_h$, multiplying
both sides by $\exp(-iN_0\varphi_e)$, and integrating over
$\varphi_e$, we obtain
\begin{equation}
\label{eq-integral}
\int_0^{2\pi}d\varphi_e\Psi(\varphi_e,\varphi_e)e^{-iN_0\varphi_e}
=\sum_N A_{N,N_0-N}.
\end{equation}
Substituting (\ref{eq-simplified}) into (\ref{eq-integral}) we arrive
at the desired condition
\begin{equation}
\label{eq-condition}
1+V_0\sum_N\frac{1}{E_N^{(e)}+E_{N_0-N}^{(h)} - \Delta_{N_0}}=0.
\end{equation}
For each integer $N_0$ the solutions of Eq.\ (\ref{eq-condition}) form
a discrete set, $\Delta_{N_0}^m$. The corresponding (non-normalized)
wave functions have the form
\begin{equation}
\label{eq-wf}
\Psi_{N_0}^m\propto
e^{iN_0\varphi_h}
\sum_N\frac{e^{iN(\varphi_e-\varphi_h)}}{E_N^{(e)}+E_{N_0-N}^{(h)}
  - \Delta_{N_0}^m}.
\end{equation}
The exponential factor in front of the sum insures that in the dipole
approximation only the excitons with $N_0=0$ can be created by light.
The frequency dependence of the exciton absorption, $\alpha(\omega)$,
can be presented as
\begin{equation}
\label{eq-spectral}
\alpha(\omega)\propto\sum_mF_m\delta(\hbar\omega-E_g-\Delta_0^m),
\end{equation}
where $E_g$ is the band-gap of the material of the ring; the
coefficients $F_m$ stand for the oscillator strengths of the
corresponding transitions. A general expression for $F_m$ through the
eigenfunction, $\Psi_0^m$, of the excitonic state reads
\begin{equation}
\label{eq-strength}
F_m=\frac{|\int_0^{2\pi}d\varphi_e\int_0^{2\pi}d\varphi_h
  \Psi_0^m(\varphi_e,\varphi_h)
  \delta(\varphi_e-\varphi_h)|^2}
  {\int_0^{2\pi}d\varphi_e\int_0^{2\pi}d\varphi_h
  |\Psi_0^m(\varphi_e,\varphi_h)|^2}.
\end{equation}
Upon substituting Eq.\ (\ref{eq-wf}) into Eq.\ (\ref{eq-strength}) and
making use of Eq.\ (\ref{eq-condition}), we obtain
\begin{equation}
\label{eq-new}
F_m=\Biggl[V_0^2\sum_N\frac{1}{(E_N^{(e)}+E_{-N}^{(h)} -
  \Delta_0^m)^2}\Biggr]^{-1}.
\end{equation}
The latter expression can be presented in a more compact form by
introducing the rate of change of the exciton energy with the
interaction parameter $V_0$.  Indeed, taking the differential of Eq.\
(\ref{eq-condition}), yields
\begin{equation}
\label{eq-last}
F_m=-\frac{\partial\Delta_0^m}{\partial V_0}.
\end{equation}

We note that the summation in Eq.\ (\ref{eq-condition}) can be carried
out in a closed form by using the identity
\begin{equation}
\label{eq-identity}
\sum_{N=-\infty}^{\infty}\frac{1}{(\pi N-a_1)(\pi N-a_2)}=
\frac{1}{(a_1-a_2)}\Biggl(\frac{1}{\tan a_2}-\frac{1}{\tan
  a_1}\Biggr).
\end{equation}
For the most interesting case $N_0=0$ the parameters $a_1$, $a_2$ are
equal to
\begin{equation}
\label{eq-12}
a_{1,2}=-\pi\Biggl[\frac{\Phi}{\Phi_0}\pm
\Bigl(\frac{\Delta_0^m}{\varepsilon_0} \Bigr)^{1/2}\Biggr],
\end{equation}
where
\begin{equation}
\label{eq-eps}
\varepsilon_0=\frac{\hbar^2}{2\rho^2}\Bigl(\frac{1}{m_e}+\frac{1}{m_h}\Bigr)=
\frac{\hbar^2}{2\mu \rho^2},
\end{equation}
and $\mu=m_em_h/(m_e+m_h)$ denotes the reduced mass of electron and
hole. Then the equation (\ref{eq-condition}) for the exciton energies
takes the form
\begin{equation}
\label{eq-form}
\Biggl(\frac{\Delta_0^m}{\varepsilon_0}\Biggr)^{1/2}=-\Biggl(\frac{\pi
  V_0}{\varepsilon_0}\Biggr)\frac{\sin
  \Bigl(2\pi(\Delta_0^m/\varepsilon_0)^{1/2}\Bigr)} {\cos
  \Bigl(2\pi(\Delta_0^m/\varepsilon_0)^{1/2}\Bigr)-\cos\Bigl(2\pi(\Phi/\Phi_0)
  \Bigr)}.
\end{equation}
This equation is our main result. It is seen from Eq.\ (\ref{eq-form})
that the structure of the excitonic spectrum is determined by a
dimensionless ratio $|V_0|/\varepsilon_0$.  From the definition
(\ref{eq-not}) it follows that, with increasing the radius $\rho$ of
the ring, $V_0$ falls off as $1/\rho$. Thus, $|V_0|/\varepsilon_0$ is
proportional to $\rho$. In the limit of large $\rho$, when $|V_0|\gg
\varepsilon_0$, the spectrum can be found analytically. The ground
state corresponds to negative energy and is given by
\begin{equation}
\label{eq-correction}
\Delta_0^0=-\frac{\pi^2V_0^2}{\varepsilon_0}\Biggl[1+4\cos\Bigl(\frac{2\pi\Phi}
{\Phi_0}\Bigr)\exp\Bigl(-\frac{2\pi^2|V_0|}{\varepsilon_0}\Bigr)\Biggr].
\end{equation}
We note that the prefactor $\pi^2V_0^2/\varepsilon_0$ is independent
of $\rho$.  It is equal to the binding energy of an exciton on a
straight line.  It is easy to see that in the limit under
consideration we have $|\Delta_0^0|\gg|V_0|\gg\varepsilon_0$.

The second term in the brackets of Eq.\ (\ref{eq-correction})
describes the AB effect for the exciton. In the limit of large $\rho$
its magnitude is exponentially small. The physical meaning of the
exponential prefactor can be understood after rewriting it in the form
$\exp(-2\pi\rho\gamma)$, where
$\gamma=\pi|V_0|\Bigl(2\mu/\hbar^2\varepsilon_0\Bigr)^{1/2}$ is the
inverse decay length of the wave function of the internal motion of
electron and hole in the limit $\rho\rightarrow\infty$. Thus, the
magnitude of the AB effect in the limit of large $\rho$ represents the
amplitude for bound electron and hole to tunnel in the opposite
directions and meet each other ``on the opposite side of the ring''
(opposite with respect to the point where they were created by a
photon).  This qualitative consideration allows to specify the
condition that the interaction potential is short-ranged. Namely, for
Eq.\ (\ref{eq-correction}) to apply, the radius of potential should be
much smaller than $\gamma^{-1}$. It is also clear from the above
consideration that, within a prefactor, the magnitude of the AB effect
is given by $\exp(-2\pi\rho\gamma)$ for arbitrary attractive
potential, as long as the decay length $\gamma^{-1}$ is smaller than
the perimeter of the ring. In Figs.\ \ref{fig-dbe-x} and
\ref{fig-be-x} we plot the numerical solution of Eq.\ (\ref{eq-form})
for various values of $\Phi$ together with the asymptotic solution
(\ref{eq-correction}) valid in the limit of large $\gamma\rho$. We see
that the maximum possible change in exciton energy by threading the
ring with a flux $\Phi_0/2$ is $25\%$ of the size-quantization energy
$\varepsilon_0$.  The asymptotic expression of (\ref{eq-correction})
is good down to $\gamma\rho\approx \pi^{-1}$.  In Fig.\ 
\ref{fig-be-phi-x}, we show the variation of the exciton energy with
$\Phi$ within one period. As expected, the AB oscillations are close
to sinusoidal for large values of $2\pi\gamma\rho$, whereas for
$2\pi\gamma\rho = 1$, unharmonicity is already quite pronounced.  The
increase of the exciton energy as the flux is switched on has a simple
physical interpretation. If the single-electron energy
(\ref{eq-energies}) {\em grows} with $\Phi$ then the single-hole
energy is {\em reduced} with $\Phi$ and vice versa. This suppresses
the electron-hole binding.  Fig.\ \ref{fig-be-phi-x} illustrates how
the amplitudes of the AB oscillations decrease with increasing ring
perimeter $2\pi \gamma\rho$ as described by Eq.\ 
(\ref{eq-correction}).  The AB oscillations in the oscillator strength
are plotted in Fig.\ \ref{fig-os-phi-x}. As expected, the shift is
most pronounced for $\Phi= \Phi_0/2$, and the relative magnitude is
nearly $80\%$ for the smallest value of $2\pi \gamma\rho$. For larger
values of $2\pi\gamma\rho$, the oscillations in $F_0(\Phi)$ become
increasingly sinusoidal as can be seen by differentiating Eq.\ 
(\ref{eq-correction}) with respect to $V_0$.

In the consideration above we assumed the width of the ring to be
zero. In fact, if the width is finite but smaller than the radius of
the exciton, $\gamma^{-1}$, it can be taken into account in a similar
fashion as in \cite{WenF95} by adding $\hbar^2 \pi^2 / 2 m_e W^2$ and
$\hbar^2 \pi^2 / 2 m_h W^2$ to the single-electron and single-hole
energies (\ref{eq-energies}), respectively. Here, $W$ stands for the
width of the ring and a hard-wall confinement in the radial direction
is assumed. This would leave the AB oscillations unchanged. In the
opposite case $W \gg \gamma^{-1}$ the oscillations are suppressed. The
precise form of the suppression factor as a function of
$(W\gamma)^{-1}$ is unknown and depends on the details of the
confinement.

Let us briefly address the excited states of the exciton corresponding
to $m>0$. In the limit $|V_0| \gg \varepsilon_0$ for the energies with
numbers $m < |V_0|/\varepsilon_0$ we get from Eq.\ (\ref{eq-form})
\begin{equation}
\label{eq-excited}
\Delta_0^m = \frac{\varepsilon_0}{4} \Bigl[ m^2 +
(-1)^m(m+\frac{1}{2}) \frac{\varepsilon_0}{\pi^2 V_0}
\cos\Bigl(\frac{2\pi\Phi}{\Phi_0}\Bigr) \Bigr].
\end{equation}
In contrast to the ground state as in (\ref{eq-correction}) the AB
contribution to the energy $\Delta_0^m$ is not exponentially small.
Still the AB term is small (in parameter $\varepsilon_{0} / |V_0| \ll
1$) compared to the level spacing at $\Phi=0$.

An alternative way to derive Eq.\ (\ref{eq-form}) is to follow the
Bethe ansatz approach.\cite{Mat93} The intimate relation between Eq.\
(\ref{eq-form}) and a Bethe ansatz equation becomes most apparent in
the absence of magnetic flux, $\Phi=0$, when (\ref{eq-form}) can be
rewritten as
\begin{equation}
\label{eq-bethe}
2\pi\rho k_m= 2 \pi m + 2 \arctan \Bigl(\frac{\rho k_m}{c}\Bigr),
\end{equation}
where $k_m=(2\Delta_0^m\mu)^{1/2}/\hbar$ is the wave vector and $c= 2
\pi \mu V_0 \rho^2/\hbar^2$ parameterizes the strength of the
attraction analogously to the well-known $\delta$-function gas
.\cite{LieL63,Lie63,Mcg64} At finite flux, the structure of the Bethe
ansatz equations will be very similar to the equations for a 1D
Hubbard model \cite{LieW68} in the presence of a spin flux coupling to
the spin-up and spin-down degrees of freedom of the electrons
.\cite{RomS94b,RomP95} We emphasize that in such discrete models the
periodicity will also be influenced by whether the number of sites in
the ring is even or odd \cite{WuM91} in addition to the continuous
situation considered in the present manuscript.

First experimental studies of the AB effect were carried out on
metallic rings.\cite{Was91} The next generation of rings were based on
GaAs/AlGaAs hetereostructures as in Refs.\ \onlinecite{MaiCB93} and
\onlinecite{YacHMS95} and had a circumference of $\sim 6000$nm and
$3000$nm, respectively. For such rings the magnitude of the excitonic
AB oscillations will be very small. However, quite recently much more
compact ring-shaped dots of InAs in GaAs with a circumference of $\sim
250$nm were demonstrated to exist.\cite{LorLFK99,LorLGK99} This was
achieved by modification of a standard growth procedure
\cite{LeoKRD93} used for the fabrication of arrays of self-assembled
InAs quantum dots in GaAs. Recent light absorption experiments on
nano-rings reveal an excitonic structure.\cite{PetWLK00} However, it
is much more advantageous to search for the AB oscillations proposed
in the present paper not in absorption, but in luminescence studies.
This is because near-field techniques developed in the last decade
allow to "see" a single quantum dot and thus avoid the inhomogeneous
broadening. This technique was applied to many structures containing
ensembles of quantum dots ({\em e.g.},
GaAs/AlGaAs,\cite{BruBAW92,BruABT94,BruABT94b,HesBHP94,ZreBHA94,BocRFH96,GamSSK96,GamSSK96b,GamBSK97,WegSAR97,BonEGP98}
ZnSe\cite{KumWBF98}).  In particular, extremely narrow and temperature
insensitive (up to $50$K) luminescence lines from a single InAs
quantum dot in GaAs were recorded in Refs.\ 
\onlinecite{MarGIB94,GruCLB95,DekGES98}.

In conclusion, we have demonstrated the AB oscillations for a {\em
  neutral} object. This constitutes the main qualitative difference
between our paper and previous considerations \cite{WenFC95} for two
interacting {\em electrons} on a ring. Lastly, we note that the
possibility of the related effect of Aharonov-Casher oscillations for
an exciton was considered previously in Ref.\ \onlinecite{KriK94}. The
underlying physics in Ref.\ \onlinecite{KriK94} is that even a {\em
  zero-size} exciton having zero charge can still have a finite {\em
  magnetic moment}.

Upon completion of this work we have been made aware of Ref.\ 
\onlinecite{Cha95} in which the underlying physics of the AB
oscillations of excitonic levels was uncovered.  Although the
analytical approach employed in Ref.\ \onlinecite{Cha95} is different
from ours, the result obtained for the ground state energy is similar
to Eq.\ (\ref{eq-correction}).

\acknowledgements
This work was supported by the NSF-DAAD collaborative research grant
INT-9815194. MER was supported in part by NSF grant DMR 9732820.  RAR
also gratefully acknowledges the support of DFG under
Sonderforschungsbereich~393. We are grateful to M.\ B\"{u}ttiker, A.\ 
Lorke, T.\ V.\ Shahbazyan, R.\ Warburton, and J.\ Worlock for useful
discussions. We thank A.\ V.\ Chaplik for pointing out Ref.\ 
\onlinecite{Cha95} to us.




\newpage \newcommand{\figwidth}{0.9\columnwidth}

\begin{figure}
\centerline{\psfig{file=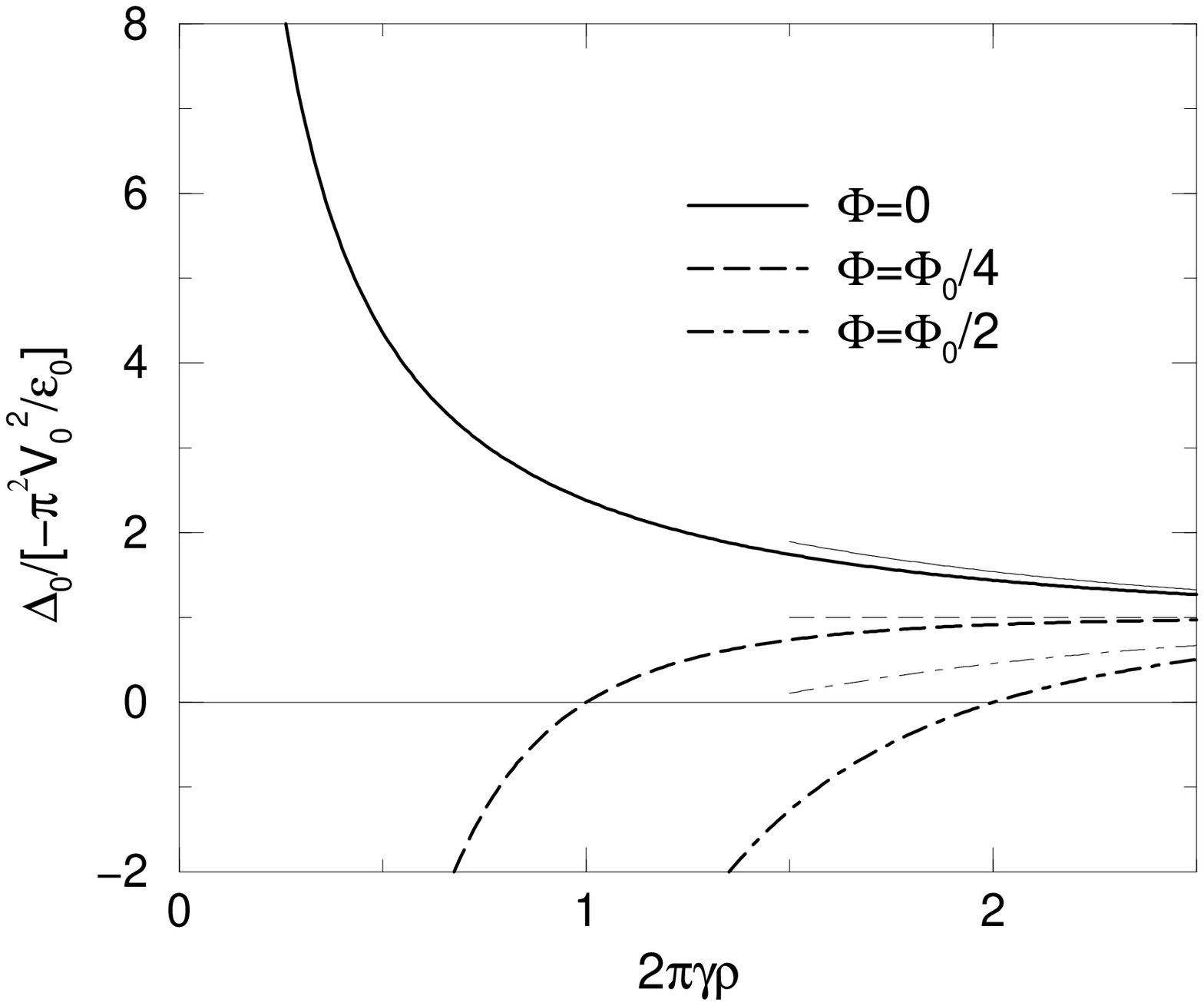,width=\figwidth}}
\caption{\label{fig-dbe-x}
  The dimensionless binding energy (in units of
  $\pi^2|V_0|^2/\varepsilon_0$) at flux $\Phi=0$ (solid lines),
  $\Phi_0/4$ (dashed line), and $\Phi_0/2$ (dot-dashed line) through
  the ring plotted versus the dimensionless perimeter of the ring
  $2\pi\gamma\rho$. The thick and thin lines represent the exact
  solution of Eq.\ (\protect \ref{eq-form}) and the asymptotic result
  of Eq.\ (\protect\ref{eq-correction}), respectively. }
\end{figure}

\begin{figure}
  \centerline{\psfig{file=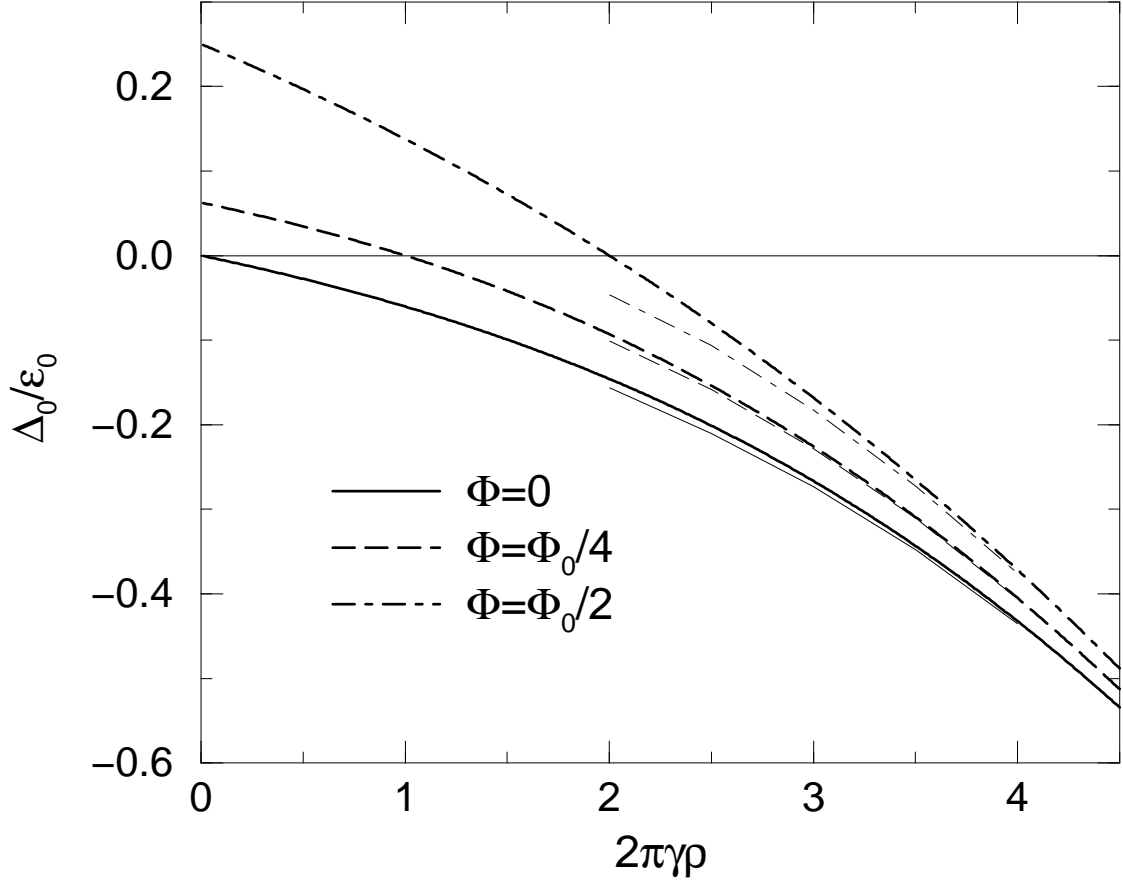,width=\figwidth}}
\caption{\label{fig-be-x}
  The exciton energy $\Delta_0/\varepsilon_0$ at flux $\Phi=0$ (solid
  lines), $\Phi_0/4$ (dashed line), and $\Phi_0/2$ (dot-dashed line)
  through the ring are plotted versus the dimensionless perimeter of
  the ring $2\pi\gamma\rho$. The thick and thin lines represent the
  exact and the asymptotic result as in Fig.\ \ref{fig-dbe-x},
  respectively.}
\end{figure}

\begin{figure}
  \centerline{\psfig{file=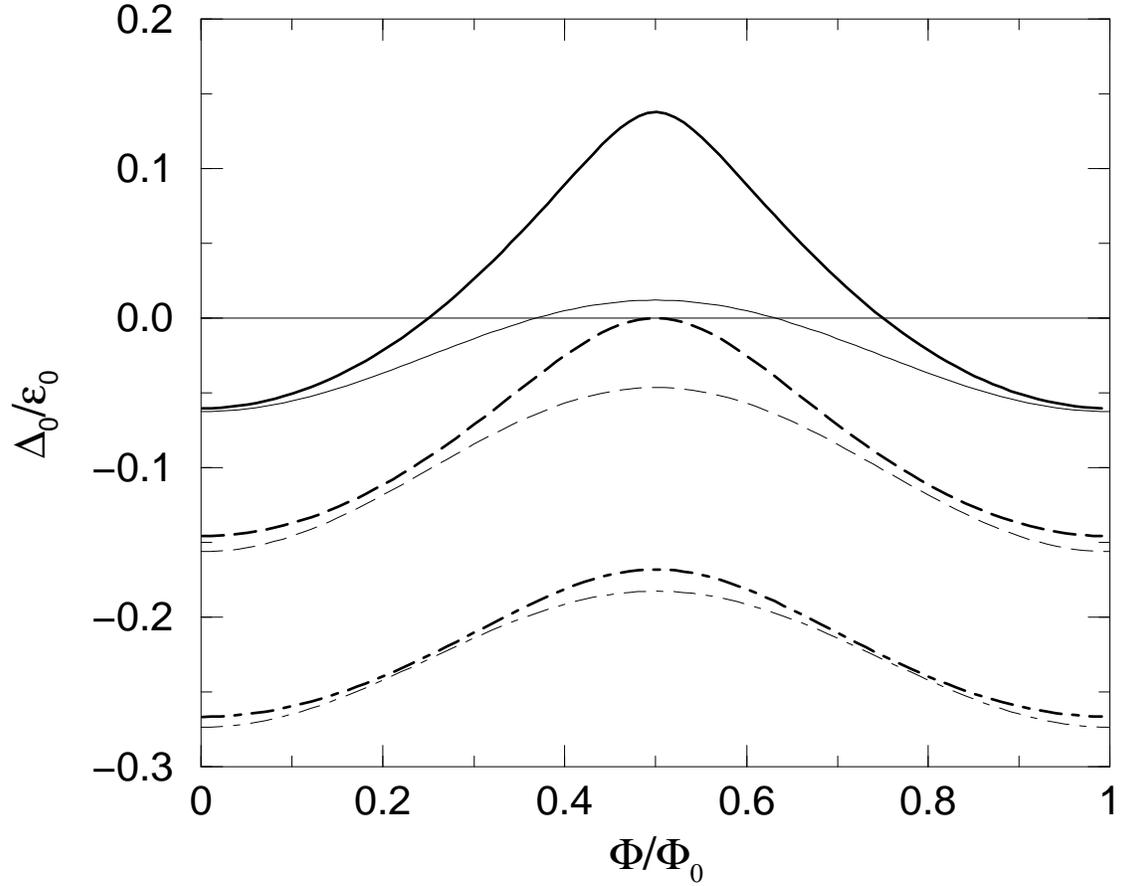,width=\figwidth}}
\caption{\label{fig-be-phi-x}
  The Aharonov-Bohm oscillations of the exciton energy is shown for
  three values of the dimensionless ring perimeter $2\pi\gamma\rho =
  1$ (solid lines), $2$ (dashed lines) and $3$ (dot-dashed lines). As
  in Fig.\ \protect\ref{fig-dbe-x}, the thick and thin lines are drawn
  from Eq.\ (\protect \ref{eq-form}) and Eq.\
  (\protect\ref{eq-correction}), respectively.}
\end{figure}

%

\begin{figure}
  \centerline{\psfig{file=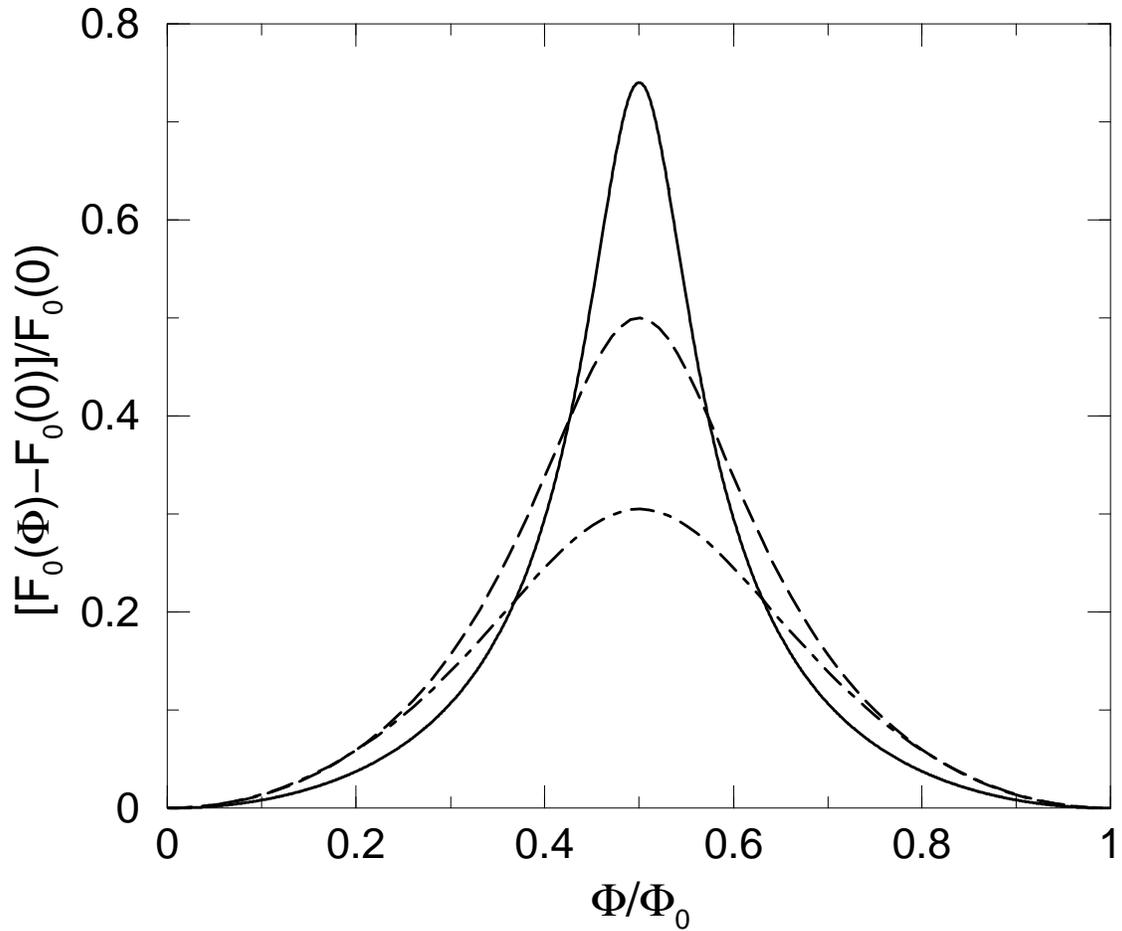,width=\figwidth}}
\caption{\label{fig-os-phi-x}
  The Aharonov-Bohm oscillations of the oscillator strength for the
  three values of the dimensionless ring perimeter $2\pi\gamma\rho =
  1$ (solid line), $2$ (dashed line) and $3$ (dot-dashed line).}
\end{figure}


\end{document}